\documentclass[twocolumn,aps,prb,superscriptaddress]{revtex4}

\usepackage{ae}
\usepackage[T1]{fontenc}
\usepackage[ansinew]{inputenc}
\usepackage{amsmath}
\usepackage{amssymb}
\usepackage{graphicx}
\usepackage{color}
\usepackage[colorlinks]{hyperref}

\begin{document}

\title{Magnetomotive Instability and Generation of Mechanical Vibrations in Suspended Semiconducting Carbon Nanotubes}
\author{A. Nordenfelt}
\affiliation{Department of Physics, University of Gothenburg, SE-412
96 G{\" o}teborg, Sweden}
\author{Y. Tarakanov}
\affiliation{Department of Applied Physics, Chalmers University of Technology, SE-412
96 G{\" o}teborg, Sweden}
\author{L. Y. Gorelik}
\affiliation{Department of Applied Physics, Chalmers University of Technology, SE-412
96 G{\" o}teborg, Sweden}
\author{R. I. Shekhter}
\affiliation{Department of Physics, University of Gothenburg, SE-412
96 G{\" o}teborg, Sweden}
\author{M. Jonson}
\affiliation{Department of Physics, University of Gothenburg, SE-412
96 G{\" o}teborg, Sweden}
\affiliation{SUPA, Department of Physics, Heriot-Watt University, Edinburgh EH14 4AS,
Scotland, UK}
\affiliation{Division of Quantum Phases
and Devices, School of Physics, Konkuk University, Seoul 143-701, Korea}
\date{\today}

\begin{abstract}
We have theoretically investigated the electromechanical properties of 
 a freely suspended carbon nanotube 
that is connected to a constant-current source and subjected to an external magnetic field. 
We show that 
self-excitation of mechanical vibrations of the nanotube can occur 
 if the magnetic field $H$ exceeds a dissipation-dependent critical value $H_{c}$, which we find to be of the order of 
 10-100 mT for realistic parameters. The instability 
 develops into a stationary regime characterized by time periodic oscillations in the fundamental bending mode amplitude. We find that for 
 nanotubes with large quality factors and a 
 magnetic-field strength just above $H_{c}$ the frequency of the stationary vibrations is very close to the eigenfrequency of the fundamental mode. We also demonstrate that the magnetic field dependence of the time averaged voltage drop across the nanotube has a singularity at $H=H_{c}$. We discuss the possibility of using this phenomenon for the detection of nanotube vibrations.
\end{abstract}

\maketitle

Nanoelectromechanical systems (NEMS) 
 containing a suspended carbon nanotube (CNT) vibrating at radio frequencies (RF) have received increasing attention recently. 
Advantages of CNT mechanical resonators include their high resonance frequencies --- up to the GHz range ---- their low dissipative losses\cite{zant2} and the possibility to tune the resonance frequency by adjusting the tension in the tube
\cite{sazonova, zant1}.
CNT based NEMS devices have already shown a great potential for a plethora of technological applications including mass sensing\cite{bachtold, Zettl-1} and tunable high frequency electronics \cite{sazonova, zant1, isacsson, eriksson}. However, most of the devices which have been realized thus far are passive resonators which perform frequency filtering of the incoming RF-signal\cite{sazonova, zant1, isacsson, eriksson}. 
Here we propose an active oscillator based on an current biased doubly-clamped CNT for which the conductance depends monotonically on the nanotube deflection. Such a deflection sensitive resistance has been demonstrated for a semiconducting single-walled CNT suspended over a gate electrode\cite{sazonova, zant1, yury}. The active feedback is provided by a Lorentz force induced by a constant magnetic field directed perpendicular to the direction of the current and parallel to the gate electrode. We show that 
, by applying a constant external current in a sufficiently high magnetic field, we obtain mechanical instability and self-sustained mechanical oscillations at a frequency close to the mechanical resonance frequency. Furthermore, we show that the mechanical instability results in oscillations of the voltage drop across the nanotube accompanied by a deviation of the time averaged voltage from its static time-independent value.

\begin{figure}
\includegraphics[width=\columnwidth]{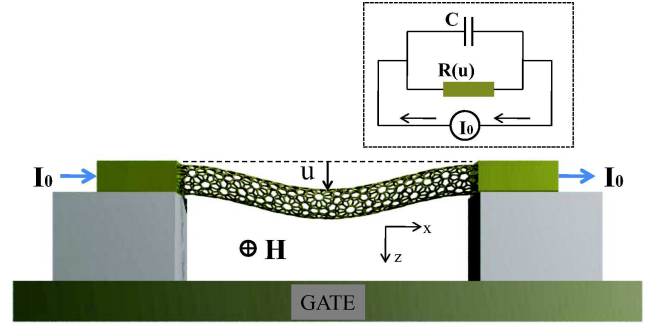}
\caption{
Sketch of the proposed active oscillator device. A semiconducting carbon nanotube is suspended over a gate electrode and
connected to an external dc current source. An external magnetic field, applied perpendicular to the direction of the current, gives rise to a Lorentz force that deflects the tube towards the gate, which affects the resistance and provides a feedback mechanism that for large enough magnetic fields leads to self-sustained nanotube oscillations (see text). The inset shows an equivalent electric circuit}\label{setup}
\end{figure}

The proposed oscillator device is shown in Figure $\ref{setup}$.
A semiconducting carbon nanotube is suspended over a gate electrode and subject to a homogeneous magnetic field $H$ perpendicular to the nanotube and parallel to the gate electrode. The gate-nanotube separation depends on the deflection $z(t,x)$ of the tube from its equilibrium straight configuration ($x$ is the coordinate along the nanotube).  
We will assume that the nanotube mechanics is completely characterized by the amplitude of the fundamental bending mode $u(t)$, and consequently 
let the
time dependence of the mechanical deflection 
have
the form $z(t,x) = u(t)\varphi_{0}(x)$, where $\varphi_{0}(x)$ is the normalized profile of the fundamental mode\cite{LL}. This assumption captures the essence of the physics behind the phenomena to be considered below, 
for which the dependence of 
the mutual capacitance $C_{G}(u)$ between nanotube and gate on the amplitude of the fundamental mode is central. 
This is because
the concentration of charge carriers in the semiconducting nanotube $\rho(C_G(u),V_{g})$ is a function of both the gate voltage $V_G$ and the mutual gate capacitance $C_G(u)$, which means that ultimately the nanotube resistance $R(\rho(C_G(u),V_G))$ will depend on the nanotube deflection. In the experiments of Sazonova et al\cite{sazonova}, this effect was utilized for the detection of resonant mechanical excitations of the nanotube.

When an 
external current-source 
feeds a current $I_{0}$ to one of the leads,
a Lorentz force proportional to the current through the nanotube and the magnetic field causes a deflection of the CNT. For displacements on a scale much smaller than the length of the wire, its curvature is negligible and consequently one can consider the Lorentz force to be the same at every point. Since the force is almost uniform over the wire, 
it is mainly the fundamental bending mode that will be affected, which provides another justification for our assumption that all other modes can be neglected. 
Assuming the nanowire to be an elastic beam whose motion can be described by linear continuum mechanics, we arrive at the following set of equations\cite{comment1} for the time evolution of the voltage drop $V(t)$ across the nanotube and the amplitude of the fundamental mode $u(t)$:
\begin{align}\label{main}
  &m\ddot{u}(t) + \gamma \dot{u}(t) +\kappa u(t) = LHI \\
  &C\dot{V}(t)= I_{0} - I,\,\,  I =V/R(u(t))\,.  \nonumber
\end{align}
Here $m$ and $L$ are the effective mass and length of the suspended part of the nanotube, $\kappa$ and $\gamma$ are the effective spring and damping constants, and $C$ is the capacitance of the junction. 

The system of equations (\ref{main}) has a time independent solution given by
\begin{align}\label{IVa}
&u(t)=u_{0}=LHI_{0}/\kappa \\
&V(t) = V_{0} = R\left( \frac{LHI_{0}}{\kappa}\right)I_{0}.\label{IVb}
\end{align}
A linear stability analysis of this solution yields a secular equation for the Lyapunov exponents $\lambda$ of the form
\begin{equation}\label{secular}
P(\lambda,\beta) = (\lambda^2 + Q^{-1}\omega_{0} \lambda + \omega_{0}^{2})(\lambda +\omega_{R} ) - \beta \omega_{0}^{2}\lambda=0\,,
\end{equation}
 where $\omega_{0}=\sqrt{\kappa/m}$ is the eigenfrequency of the fundamental mode, $\omega_{R}=1/R(u_{0})C$, and $Q=\sqrt{\kappa m}/\gamma$ is the quality factor characterizing the damping of the mechanical vibrations due to its interaction with the thermodynamic environment.
The parameter
\begin{equation}\label{betta}
\beta= \frac{LHI_{0}}{\kappa\ell_R},\;\;   \hbox{where} \;\; \ell_R= -\frac{R(u_0)}{R_u'(u_0)}\,,
\end{equation}
is proportional to the Lorentz force and will be referred to as the electromechanical coupling parameter of the system.
It is worth to note that since $u_0$, and consequently $\ell_R$, is a function of the magnetic field and the external current, $\beta$ is not necessarily 
directly proportional to the externally adjustable parameters $H$ or $I_{0}$. How much this affects the possibility to control the electromechanical coupling through the magnetic field depends on the specific geometry and the values of other parameters such as $V_g$. It turns out that there is a simple relation between the value of the coupling parameter
$\beta$  and the sign of the differential resistance in the static regime. Indeed, from the time independent solution (\ref{IVb}) we obtain the relation
\begin{equation}\label{IV}
\frac{dV_{0}}{dI_{0}}=R(u_{0})(1-\beta),
\end{equation}
from which we see that the differential resistance in the time independent regime is positive if $\beta < 1$, while for $\beta > 1$ it is negative.

For a high quality factor, $Q \gg 1$, and weak electromechanical coupling, $\beta\ll 1$, the (approximate) solutions of equation (\ref{secular}) --- together with a third root which is always real and negative --- are
\begin{equation}\label{roots}
\lambda_{1,2}=\frac{\omega_{0}}{2}\left(\frac{\omega_{0}\omega_{R}}{\omega_{0}^{2}+\omega_{0}^{2}}\beta -\frac{1}{Q}\right)\pm i\omega_{0}\left( 1-\frac{\beta}{2} \frac{\omega_{0}^{2}\beta}{\omega_{R}^{2}+\omega_{0}^{2}}\right)\,.
\end{equation}
From equation ($\ref{roots}$) follows that the two Lyapunov exponents $\lambda_{1,2}$ have a positive real part when
\begin{equation}\label{inst}
\beta>\beta_{c}=\frac{1}{Q} \left( \frac{\omega_{0}}{\omega_{R}}+\frac{\omega_{R}}{\omega_{0}}\right).
\end{equation}
This means that for large quality factors, $Q \gg \max(\omega_{0}/\omega_{R},\omega_{R}/\omega_{0})$, 
the regime of time independent charge transport through a static
nanotube becomes unstable with respect to self-excitation 
of mechanical vibrations, if the coupling parameter $\beta$ exceeds the critical value $\beta_c$. The natural way to reach this self-excitation regime is to increase the magnetic field $H$, since increasing $I_{0}$ could lead to overheating. Hence we choose to characterize a particular setup by the critical magnetic field $H_c$ above which the system will be unstable. From equation ($\ref{inst}$) follows that for large quality factors and $\omega_0 = \omega_R$ the critical magnetic field is\cite{comment2}
$$H_{c}=2\kappa\ell_{R}/QLI_{0}.$$ If we estimate the characteristic length $\ell_{R}$ as the distance between nanotube and gate electrode and take 
$\ell_{R}\approx$ 0.1-1 $\mu$m, $L \approx$ 1 $\mu$m, $\kappa \approx 10^{-5}$ N/m, $Q \approx 1000$  and $I_{0} \approx$ 0.1 $\mu$A, we obtain a critical magnetic field of the order $H_c \sim$ 10-100 mT.

In the above section we analyzed the situation when the mechanical friction is very small. It turns out that the time independent regime becomes unstable for large enough electromechanical coupling even if the mechanical subsystem is overdamped, that is if $Q \ll 1$. Indeed, the system becomes unstable when the real part of the Lyapunov exponents changes sign from negative to positive. It means that at a certain critical point $\beta=\beta_{c}$ the secular equation (\ref{secular}) has two purely imaginary roots $\lambda_{1,2}=\pm i\omega_{c}$. These are obtained from the 
two equations
\begin{align}\label{crit}
  &\mathrm{Re} P(i\omega_{c},\beta_{c})=\omega_{R}\omega^{2}_{0}-\omega^{2}_{c}(\omega_{R}+Q^{-1}\omega_{0})=0\\
  &\mathrm{Im }P(i\omega_{c},\beta_{c})= -\omega_{c}(\omega_{c}^{2}+Q^{-1}\omega_{0}\omega_{R}-\beta_{c})=0\,, \nonumber
\end{align}
whose general solution is 
\begin{align}\label{crit2}
  &\beta_c = \frac{1}{Q}\frac{\omega_{0}\omega_{R} +Q(\omega_{0}^{2} + \omega_{R}^{2})}{\omega_{0}(\omega_{0} + Q\omega_{R})}\\
  &\omega_{c}=\omega_{0}\sqrt{\frac{Q\omega_{R}}{\omega_{0}+Q\omega_{R}}}. \nonumber
\end{align}
Important to note is that $\beta_c$ is always positive. In particular, from this follows that the instability can only occur when the direction of the magnetic field is such that it deflects the nanotube towards decreasing resistance. In our geometry it means that the Lorentz force must be directed towards the gate electrode.

From equations (\ref{crit2}) it follows that for small quality factors, $Q \ll \max (1,\omega_{0}/\omega_{R})$,  and when the coupling parameter $\beta$ has just overcome the critical value $\beta_{c}\cong \omega_{R}/(Q\omega_{0})$, the static regime becomes unstable with respect to oscillations of angular frequency $\omega_{c}\approx \sqrt{\omega_{0}\omega_{R}Q}$, which is always smaller than the mechanical resonance frequency. It is worth noting that, for an overdamped mechanical system the critical value of the coupling parameter is bigger than unity, and as discussed previously this implies
that the instability in this regime could be considered to be caused by a negative differential resistance. Estimating the critical magnetic field with the same parameters as before, but now with $Q=0.1$, we obtain $H_c \approx$ 100-1000 T, 
which is too large for experimental observation. Therefore, for the remaining part we will focus our attention on the low-dissipation regime, which can be achieved at least at low temperatures\cite{zant2, quantum}.

\begin{figure}
\centerline{\includegraphics[width=\columnwidth]{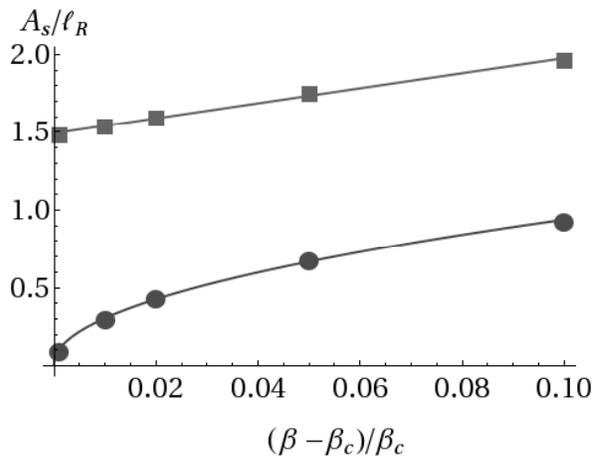}}
\caption{The above graph show the results of a number of computer simulations of the system of equations ($\ref{main}$) aimed at illustrating the regimes of soft and hard instability. In all the simulations we chose the quality factor $Q = 10^2$ and the resistance $R(u)/R_0 = (1 + e^{-2(u-u_0)/\ell_R})/2$. The stationary mechanical deflection is then equal to the coupling parameter: $u_0/\ell_R = \beta$. Initial values were $u(0) = \dot{u}(0) = V(0) = 0$. In the first series of simulations we put $\omega_R = \omega_0$ for which formula ($\ref{ab}$) predict soft instability, the solid circles mark the saturation amplitude of the mechanical oscillation for different values of the ratio $(\beta - \beta_c)/\beta_c$. The results in this first series of simulations are in very good agreement with the analytical estimates. In the second series of simulations, marked with the solid squares in the graph, we chose $\omega_R = 2\omega_0$ for which formula ($\ref{ab}$) predicts hard instability. In this regime the saturation amplitude appears to have a discontinuous jump at the threshold of instability. Two curves are fitted to the data points.}\label{softhard}
\end{figure}

The way in which the instability evolves is largely dependent on the magnitude of $\beta$.
For a coupling parameter just above the threshold, $\beta_c \lessapprox \beta \ll 1$, we may analyze the development of the instability with the Ansatz $u(t) = u_0 + A(t)\sin(\omega_{0}t)$, assuming $A(t)$ to be a slowly varying function on the scale $1/\omega_{0}$. Substituting this Anzats into Eq. (\ref{main}) and averaging over the fast oscillations \cite{Nayfeh}, we obtain an 
equation for $\dot{A}$,
\begin{equation}\label{adot}
\dot{A}(t) = a_{1}\omega_{0}A(t)\left[\left(\frac{\beta-\beta_{c}}{\beta}\right) +b_{1}\frac{A^2(t)}{(2\ell_{R})^{2}}\right]\,,
\end{equation}
where
\begin{eqnarray}\label{ab}
a_1 &=&\frac{\beta}{2} \frac{\omega_{0}\omega_{R}}{\omega_{0}^{2}+\omega_{R}^{2}} \; \;\;\;\;\;\;\;\;\;\;\;\;\;\;\;\;\;\;\;\;\;\;\;\;\nonumber\\
b_{1} &=&  \frac{4\omega_{0}^{4}-5\omega_{0}^{2}\omega_{R}^{2}+3\omega_{R}^{4}}{2(\omega_{0}^{2} + \omega_{R}^{2})(4\omega_{0}^{2} + \omega_{R}^{2})}+ \;\;\;\;\;\;\;\;\;\;\;\;\;\;\;\;\\
&& + \frac{1}{2}\left(\frac{3\omega_{R}^{2}-\omega_{0}^{2}}{\omega_{0}^{2} + \omega_{R}^{2}}\right)\frac{\partial\ell_{R}}{\partial u_{0}} -\frac{1}{2}\ell_{R}\frac{\partial^{2}\ell_{R}}{\partial u_{0}^{2}}\,. \nonumber
\end{eqnarray}
From equation (\ref{adot}) follows that for $\beta > \beta_{c}$ ($H>H_{c}$) an initially small amplitude $A(t=0)\ll 2\ell_{R}$ will 
at first 
increase exponentially in time. 
There are then two different scenarios for the further development, depending on the sign of the coefficient $b_{1}$.  If  $b_{1}<0$ and $\beta - \beta_{c}\ll \beta_{c}$ the amplitude saturates at the value
\begin{equation}\label{As}
A_{s}=2\ell_{R}\sqrt{\frac{H-H_{c}}{|b_{1}|H_{c}}}\,,
\end{equation}
which vanishes for $\beta=\beta_c$ ($H=H_c$) and corresponds to a ``soft" instability. For $b_{1}>0$ the amplitude may 
saturate at a value, which does not vanish as $\beta\to\beta_c$ (from above). This is called a ``hard" instability. The different scenarios of soft and hard instabilities are illustrated in Figure ($\ref{softhard}$).

We note in passing that in the case of a hard instability, self-excitation results in a large amplitude $\sim \ell_{R}$ of the mechanical oscillations and hence nonlinear effects in the nanotube dynamics become important. Therefore, in order to analyze this regime in detail one would have to consider nonlinear elastic mechanics. This would, however, take us too far outside the scope of the present work. 

\begin{figure}
\centerline{\includegraphics[width=\columnwidth]{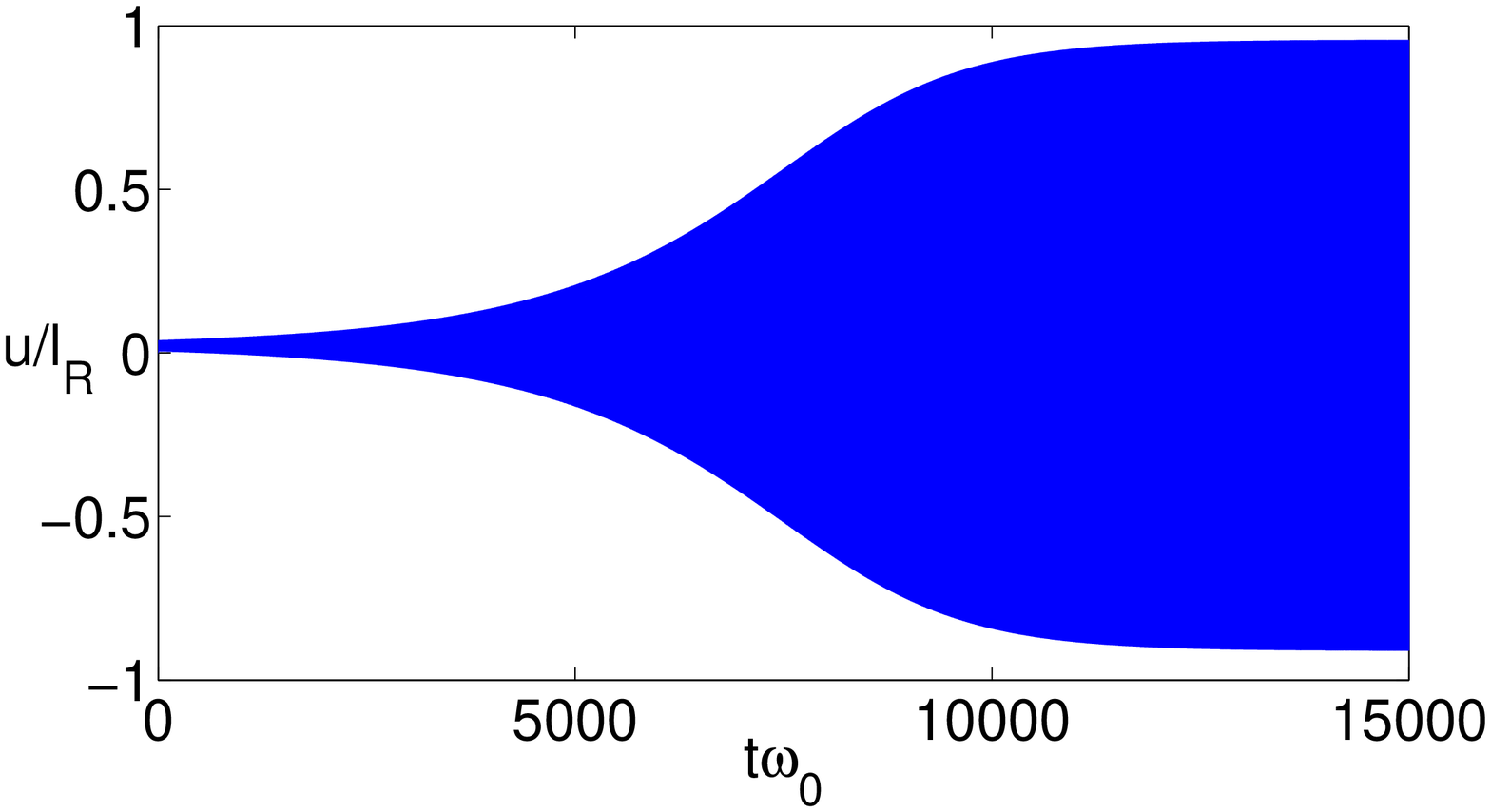}}
\centerline{\includegraphics[width=\columnwidth]{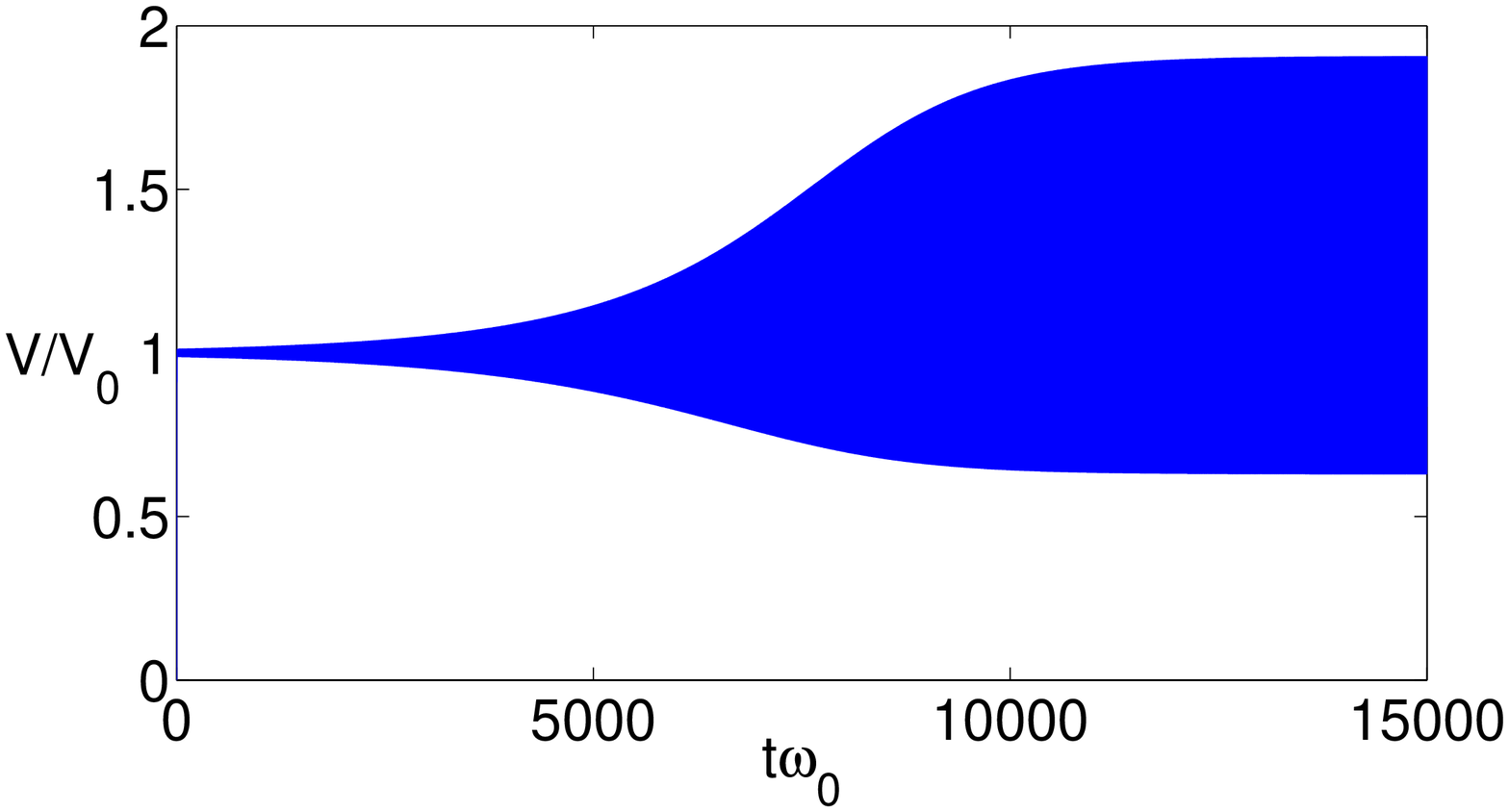}}
\caption{The above graphs show the result of a computer simulation of the system of equations ($\ref{main}$) aimed at illustrating the phenomenon of the voltage drop. For this purpose we chose the quality factor $Q = 10^2$, the RC-frequency $\omega_R = \omega_0$, the coupling parameter $\beta = 1.1\beta_c = 0.0219$ and the resistance $R(u)/R_0 = (1 + e^{-2(u-u_0)/\ell_R})/2$, where $u_0/\ell_R = \beta$. Initial values were $u(0) = \dot{u}(0) = V(0) = 0$. As can be clearly seen, the time averaged voltage deviates more and more from $V_{0}$ as the amplitude of the mechanical oscillation increases. The actual oscillations are so fast on the timescale used that they are not resolved in the graph.} \label{simulation1}
\end{figure}

The result of the instability discussed above is a stationary regime that is not only characterized by an oscillating mechanical displacement of the nanotube but also by an oscillating voltage across the trench. In the regime of high quality factors, or more precisely when $\omega_R \gg \omega_0/Q$, the oscillation frequency will be very close to the mechanical resonance frequency $\omega_0/2\pi$,  typically of the order 0.1-1 GHz. Detecting such fast voltage oscillations may pose serious experimental challenges. The nonlinearity of the system does, however, result 
in a nonzero deviation of the time-averaged voltage $\overline{V}(H)$ from the static value $V_{0}(H)$. This phenomenon is illustrated graphically in Figure ($\ref{simulation1}$). If $A_s \ll \ell_{R}$ the time-averaged voltage can be approximated to lowest order with the formula
\begin{equation}\label{vdrop}
\frac{\overline{V}(H) - V_{0}(H)}{V_{0}(H)} = \left(\frac{\partial \ell_R}{\partial u_0} + \frac{\omega_R^2 - \omega_0^2 }{\omega_R^2 + \omega_0^2}\right)\left(\frac{A_s}{2\ell_R}\right)^{2}
\end{equation}

Thus, in the case of a soft instability the saturation amplitude $A_s^{2}$ can be calculated from formula (\ref{As}) and is found to be proportional to $(H - H_c)/H_c$. Accordingly, in this case we have the following relation:
\begin{equation}
\frac{\overline{V} - V_{0}}{V_{0}} \propto \frac{H - H_c}{H_c} \theta(H - H_c)
\end{equation}
where $\theta (x)$ is the Heaviside step function. Thus, at the critical point $H = H_c$ a soft instability manifests itself as a jump in the derivative $\partial \overline{V}/\partial H$ whereas a hard instability would show up as a discontinuous jump in $\overline{V}$. Hence, the instability could be detected by measuring the average voltage drop across the suspended nanotube as a function of magnetic field.

Important to note is that for our geometry the characteristic length is of the order of the distance between the nanotube and the gate. Therefore one can expect strong nonlinear effects in the nanotube dynamics to start to dominate at an amplitude which is just a small fraction of the characteristic length. Hence the validity of formula ($\ref{adot}$) is rather limited. However, we believe that formula ($\ref{vdrop}$) could still be valid as an approximate relation between the voltage drop and the saturation amplitude. At the other extreme one could consider systems with a very short characteristic length, for example with an STM-tip positioned above the CNT (see for example Le Roy et al.\cite{leroy,Magnus}). In this case the current is determined by the probability for electrons to tunnel from the STM-tip into the CNT, and one could expect a characteristic length of about 0.1 nm.

To conclude, we expect that vibrations of a suspended carbon nanotube can be self-excited in the DC-current biased regime if a sufficiently large external magnetic field is applied perpendicular to the nanotube axis.
We support this claim with an analysis of a classical model of such a system and show that when the dissipation is sufficiently low, the static state of the nanotube becomes unstable at reasonably weak (10-100 mT)
magnetic fields. This magnetomotive instability develops into a steady state characterized by pronounced nanotube vibrations. We have also demonstrated that one effect of the 
nanotube vibrations 
is to change the magnetic field dependence of the average voltage drop across the nanotube. This phenomena may be used for the experimental detection of the magnetomotive instability. 

Other types of  active oscillators based on sustained self-oscillations of a on suspended CNT have been proposed theoretically\cite{Magnus}
and recently realized experimentally\cite{Zettl-2}. 
The approach in Ref. \onlinecite{Magnus} relies on distance dependent electron injection from a STM tip into a doubly-clamped CNT, while in Ref. \onlinecite{Zettl-2} distance dependent field emission of electrons from a singly-clamped CNT to the electrode provides a feedback mechanism that leads to controllable mechanical self-oscillations from a single DC voltage supply. The drawback of both these devices is that they require very precise geometry control for their operation and that the short distance between the CNT and the controlling electrode greatly limits the amplitude of vibration. 
The advantage of the device considered here is that the main element is a suspended-channel CNTFET --- fabricated in numerous experimental laboratories --- in which the CNT can be hundreds of nanometers from the gate electrode, and which requires lower voltages for operation.

The authors would like to thank A.M. Kadygrobov for fruitful discussions. This work was supported in part by the Swedish VR, by the EC project QNEMS (FP7-ICT-233952) and by the Korean WCU program funded by MEST through the NFR (R31-2008-000-10057-0).


\begin{thebibliography}{99}
\bibitem{comment1}
Since  the voltage bias across the trench needs to be relatively high, typically a few mV, a simple estimation shows that the magnetically induced electromotive force can be ignored.
\bibitem{comment2}
Using the estimate $\omega_0/2\pi \approx $ 0.1-1 GHz the optimal value of the capacitance, $C_{\rm opt}$, is found to lie approximately in the interval 1-100 fF.
\end{thebibliography}
\end{document}